# Spatial Indexing of Large Multidimensional Databases

I. Csabai, M. Trencséni, G. Herczegh, L. Dobos, P. Józsa, N. Purger, T. Budavári[1], A. Szalay[1]

Dept. of Physics of Complex Systems, Eötvös University, H-1518 Budapest, Pf.32, Hungary

[1]Dept. of Physics and Astronomy, The Johns Hopkins University, Baltimore, MD, USA

{csabai, trencseni, herczegh, dobos, jozsa, npurger}@complex.elte.hu, {budavari, szalay}@jhu.edu

## ABSTRACT

Scientific endeavors such as large astronomical surveys generate databases on the terabyte scale. These, usually multidimensional databases must be visualized and mined in order to find interesting objects or to extract meaningful and qualitatively new relationships. Many statistical algorithms required for these tasks run reasonably fast when operating on small sets of in-memory data, but take noticeable performance hits when operating on large databases that do not fit into memory. We utilize new software technologies to develop and evaluate fast multidimensional indexing schemes that inherently follow the underlying, highly non-uniform distribution of the data: they are layered uniform grid indices, hierarchical binary space partitioning, and sampled flat Voronoi tessellation of the data. Our working database is the 5-dimensional magnitude space of the Sloan Digital Sky Survey with more than 270 million data points, where we show that these techniques can dramatically speed up data mining operations such as finding similar objects by example, classifying objects or comparing extensive simulation sets with observations. We are also developing tools to interact with the multidimensional database and visualize the data at multiple resolutions in an adaptive manner.

## Categories and Subject Descriptors

H.2.8 [**Database Management**]: Database Application – *Data mining, Scientific database, Spatial database, Multi-dimensional database, GIS.*

## 1. INTRODUCTION

The fast development of technology, especially in computer hardware and microelectronics devices is revolutionizing most of the natural sciences with a sudden increase in the amount of measurement and simulation data [18]. We could have chosen our examples from almost any discipline, but in this work we will focus on astronomical data. In astronomy and cosmology the two major factors of revolution were the appearance of CCD chips and fast computers with cheap, large storage and processing capacity. Ten or twenty years ago cosmology was a theoretical science: there were many models, but there was no data to decide which one is valid. The major cosmological constants were determined only up to a factor of two. A decade later most of them are known with a few percent accuracy. As old questions are resolved, new ones take their place: today, astrophysics is a hot topic with problems such as dark matter and energy at the forefront. The first step in approaching these problems involves large scientific surveys which cumulate observation data on the terabyte scale.

Scientists have traditionally used large, flat files to store data and Fortran or C programs to analyze them. The architecture of current file systems makes this strategy prohibitive when the size of the data exceeds the gigabyte regime. To overcome this problem the scientific communities have started to use modern database technology, mainly the standard relational database products. One of the earliest and most elaborated examples is SkyServer [27] for the Sloan Digital Sky Survey (SDSS) [17]. SDSS will create a detailed digital map of a large portion of our Universe and store several terabytes of data in a publicly accessible archive.

Relational database systems which are robust enough for use with serious science archives are the same products used by commercial companies (SDSS uses Microsoft SQL Server). They were mainly designed to meet commercial requirements, not the scientific community's. Since SQL is not flexible enough for complex analysis, scientists only use it to filter data, and in later phases still use Fortran and C programs for complex analysis. This process usually involves unnecessary conversion of data between various representation formats since code is not running in the same place where data is stored. There are new developments that aim to change this, e.g. MS SQL Server 2005 comes with the new feature of allowing Common Language Runtime (CLR) code to run inside the database engine.

Most science data is multidimensional and continuous (coordinates, time, brightness of a star, size of a galaxy) in nature, and the models/theoretical equations use several of the variables, making them multidimensional, too. In this framework, measurements are points in the multidimensional space and scientific theories/models are hyper surfaces establishing relations between variables. From the viewpoint of the spatial database, scientific questions are hence transformed into queries which are hyper planes (linear theories) or curved surfaces (nonlinear theories). In practice these can be broken down into polyhedron queries.

Certain RDBMSs have so called spatial extensions, like Oracle Spatial or PostgreSQL's PostGIS; these are for handling 2D spatial representation of geographic data. SkyServer has a custom made 2D spatial module for handling celestial coordinates [19]. These techniques are not general enough for higher dimensional analysis. Another related technique is OLAP cubes. Although data cubes are multidimensional in nature, they deal with categorical data, not continuous variables. To differentiate our work from the widespread usage of the world *spatial* in geography sciences, we use the term *multidimensional spatial*. In the rest of this paper we will not use the whole term when it does not cause confusion.

Partitioning and indexing of multidimensional spaces is not a new topic, Oc-tree, R-tree, SS-tree, SR-tree, X-tree, TV-tree, Pyramid-tree and Kd-tree are just a few of the existing examples. Traditionally, these algorithms are designed and implemented as memory algorithms; it is not trivial to convert them for use with databases. Also, an algorithm that is optimal as a memory routine can give inferior performance compared to a theoretically suboptimal one when implemented in a database server [11].

Scientists need to intensively interact with the data to find hidden relations. They usually plot data on graphs to visually help this process, but when the number of data points exceeds the limits of the computer's memory or the dimensionality (independent variables) is high, this task becomes cumbersome.

Visualization should directly be linked with the database, and this link should be two-way: database indices should speed up visualization (database indices can directly be visualized, too) and the visual interface should transform visual manipulations into database requests. Geo-spatial tools like NASA's WorldWind and Google Earth are good examples for interfaces that dynamically change the level of detail (LOD) by getting additional data from Internet based map servers.

The central goal of this work is to develop tools for indexing, searching and visualizing multidimensional data. We work with a special astronomy data set described in Section 2, but note that multidimensional continuous data is quite widespread. We have created a prototype implementation of the kd-tree and Voronoi tessellation algorithms that run inside the database engine and are optimized for this purpose, the details of which are discussed in Section 3. Section 4 presents several scientific applications, Section 5 gives an overview of the interactive database visualization tool. In Section 6 we conclude with the discussion of future directions of the project.

## 2. MULTIDIMENSIONAL SCIENCE DATA AND COMPLEX SPATIAL QUERIES

The laws of nature involve relations between several physical quantities. To understand how stars and galaxies evolve one would need to measure their mass, temperature, chemical composition, etc. for a very large number of celestial objects. For systems where we are not sure what the basic quantities are we need to measure even more parameters.

Usually the basic parameters cannot be measured directly, but we have theories that relate them to observable parameters. For galaxies we can measure their brightness at various wavelengths of the electromagnetic spectrum from which we can calculate the underlying basic physical quantities. Each measurement is a data point in the multidimensional parameter space. The larger the dimensionality of the space, the larger the number of points we need to find to characterize the relations between the variables. In the next subsection we describe the multidimensional space of our working database.

### 2.1 The SDSS Color Space

SDSS measures more than 300 parameters for about 270M objects. The measurements are stored in a MS SQL database [27] exceeding 2 terabytes. Our study uses the 5 brightness values (called colors or magnitudes, since they are on a logarithmic scale) in five optical color bands: *u, g, r, i* and *z* for ultraviolet, green, red, infrared, and further infrared, respectively. Figure 1. shows a 2D projection of a small subset (500K) of the 270M points in our 5 dimensional space. We will call this color space since our variables are color values.

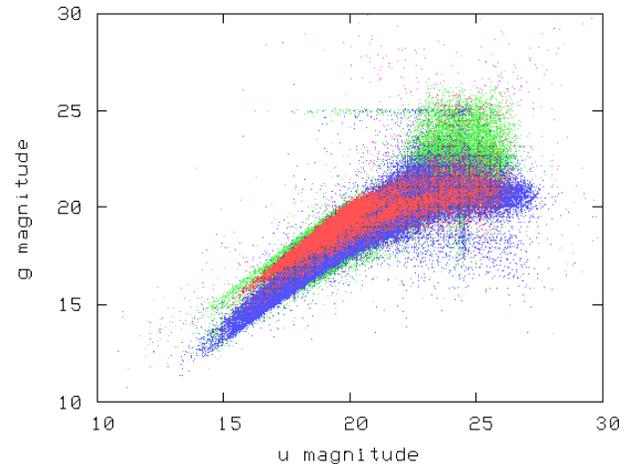

**Figure 1. A 2D projection of the 5 dimensional magnitude space of SDSS. Each point corresponds to a celestial object; green, blue and red points are stars, galaxies and quasars respectively. The distribution is highly inhomogeneous, there are outliers.**

It is evident from Figure 1. that data points do not fill the parameter space uniformly; this is typical for science data sets. There are correlations, points are clustered, they lie along (hyper)surfaces or subspaces. The correlations are sometimes exactly what we are looking for; these are relations between the variables, an expression of the underlying laws. On the other hand this uneven distribution is one of the major problems we face when indexing the data to enable efficient searches. Simple binning or quad tree like structures cannot be used in themselves efficiently; the spatial partitioning must follow the structure of the data. Also note that there are outliers, which are usually results of some measurement or calibration problem, but they can also be rare, interesting objects. These large variations in the density call for adaptive binning.

## 2.2 Typical Multidimensional Queries

The color of points in Figure 1. corresponds to the so called spectral type of the object (star, galaxy or quasar). This information is available for less than 1% of the objects in the SDSS database due to observational constraints, but classification of all objects is a crucial task for astronomy. A typical query classifies objects based on their colors, for example separates quasars from other types. To do this one should identify a few quasars with other measurements (the training set) and then draw a surface in 5D that best differentiates them from other objects. Automatic clustering, finding similar objects with drawing a convex hull around the training set or finding nearest neighbors in the color space are a few other typical problems astronomers need to solve. As an illustration, Figure 2. shows a real-life query with a complex multi-dimensional search criterion. It was selected from the more than 12 million user queries of the 2006 May log of the public SkyServer archive.

```
SELECT run, camCol, rerun, field, objID, b.ra as
input_ra, b.dec as input_dec, g.ra as sloan_ra,
g.dec as sloan_dec FROM … WHERE … and (petroMag_r
- extinction_r  <  (13.1 +  (7/3)  *  (dered_g -
dered_r) + 4 * (dered_r - dered_i) - 4 * 0.18) )
and ( (dered_r - dered_i - (dered_g - dered_r)/4 -
0.18) < 0.2) and ( (dered_r - dered_i - (dered_g -
dered_r)/4  -  0.18)  >  -0.2)  and  (  (petroMag_r -
extinction_r + 2.5 * LOG10(2 * 3.1415 * petroR50_r
*  petroR50_r))  <  24.2)  )  or  (  (petroMag_r -
extinction_r < 19.5) and ( (dered_r - dered_i -
(dered_g  -  dered_r)/4  -  0.18)  >  (0.45  -  4  *
(dered_g - dered_r)) ) and ( (dered_g - dered_r) >
(1.35  +  0.25  *  (dered_r  -  dered_i))  )  )  and  (
(petroMag_r  -  extinction_r  +  2.5  *  LOG10(2  *
3.1415 * petroR50_r * petroR50_r) ) < 23.3 ))
```

**Figure 2. SkyServer logs all user queries. To find complex spatial queries among them we have selected the top 100 ones that have magnitude values (dered_u, dered_g, dered_r, dered_i and dered_z) and '-' and '<' or '>' signs in the WHERE statement. This figure shows one of them. To save space part of the query has been left out.**

## 3. MULTIDIMENSIONAL SPATIAL INDEXING

In the geometrical sense indexing is an optimal mapping of the multidimensional space into the one dimensional index space (needed because of the sequential nature of disk I/O). We have implemented several spatial partitioning schemes: layered uniform grid indices, kd-trees [5] and Voronoi tessellation using native SQL and the .NET CLR features of MS SQL Server 2005. Our working environment is an AMD Opteron machine with 8 GB of RAM running Windows Server 2003 dedicated to SQL Server 2005 32 bit edition. Address Windowing Extensions (AWE) is enabled to manage memory over 4 GB and recovery mode was set to simple in order to avoid huge / slow log processes. We use managed code wherever possible and use wrappers to connect it to legacy C/C++ code.

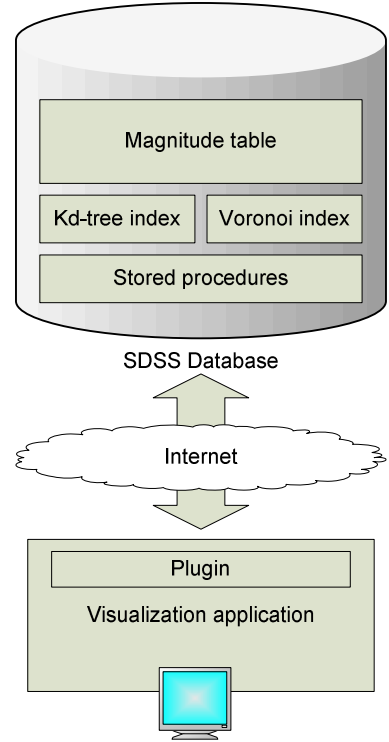

**Figure 3. System overview showing the 270M magnitude table, spatial indices, stored procedures and the adaptive visualization application.**

Throughout this discussion we assume that the database is static (like the SDSS database), e.g. no new data is inserted. Thus we will not discuss the cost of dynamic insertion or consider it a factor in assessing techniques.

## 3.1 Layered Uniform Grid

Our visualization application, to be discussed in Section 5, adaptively visualizes the first three principal components of the 270M magnitude table. The visualization application passes two parameters to the database server side: an axis aligned query box and $n$, the number of points it wants to receive from that region. The challenging requirement is to return $n$ points that follow the underlying distribution, and return them quickly. Originally, we experimented using MS SQL Server 2005's TABLESAMPLE feature and a TOP$(n)$ clause. TABLESAMPLE is a quick way to select a given $p$ percent of the database pages, and run the rest of the query on that sample. This approach worked fairly well, but it is not without problems: as the user zooms in, the client issues repeated request to the database server. Depending on the position of the query box in the global bounding box, the extents of the query box, and $n$, the number of points to be returned, $p$ must be tuned, otherwise we under sample the table and return less points, or we over sample loosing the speed advantage compared to table scan, and the TOP$(n)$ clause will return a set that does not follow the underlying distribution.

We developed a *layered uniform grid* based indexing technique that solves all these problems. First, we added a RandomID integer column to the table, which randomly orders the points (from 1 to 270M). Then we assigned the first 1024 points to the first layer, the next 8·1024 points to the second, and so on; this is stored in an integer column called Layer – thus for *RandomID < 1024, Layer = 1, for 1024 < Rid < (8 + 1) · 1024, Layer = 2,* and so on. For each layer, we created a uniform grid, where the grid resolution increases with the layer index. For the 1024 points on the first layer, we created a 2 x 2 x 2 grid, for the 8·1024 points on the next layer 4 x 4 x 4, and so on. Thus the average number of points per grid cell is constant across all layers (but the grid is not adaptive). This enables us to trivially pre-compute and store for each point on each layer which grid cell it is contained in (we number the cells in an obvious way). This is stored in an integer column called ContainedBy, e.g. for *Layer = 1, 1 < ContainedBy < 2·2·2*. In total we added three extra columns to the Magnitude table: RandomID, Layer and ContainedBy. This layered structure allows us to quickly return *n* random points independent of how large the query box is, without wasting too much time reading in useless points from disk. The visualization client passes the query box *q* and *n* to a stored procedure, which first looks at layer 1, trivially computes which of the 2 x 2 x 2 cells intersects *q*, and executes a SELECT with the conditions

```
Layer = 1 AND
ContainedBy = <cells intersecting q> AND
<point must lie in q>
```

If *r*, the number of points returned is less than *n*, go on to the next layer fetching *n – r* points, and so on. If at a given point the total number of points exceeds *n*, it halts. Since the visualization client is insensitive to receiving a little more points than it requested, extra points from the last layer are returned, too.

In practice, this approach works incredibly well. Our tests show that practically only points which are actually returned are read from disk into memory. It handles any type of query box and *n* well. An interesting feature possibility is to stream the points back to the client, i.e. when points from the first layer are available, start sending them back to the client as we fetch more points from layer 2, and so on.

### 3.2 Kd-tree Index

In multidimensional spaces the advantage of kd-trees over most other tree structures is that it applies just one cut at each level. This allows high resolution without the fast growth of the number of cells. Kd-trees can be used efficiently for outlier detection [8], nearest neighbor computation [4] and pattern search [14].

Our balanced kd-tree indexing is implemented in MS SQL Server 2005 as a set of stored procedures. Space limitations do not allow us to give all the implementation details, so we just mention one. We have experimented with many tree building schemes, always optimizing for speed. The lesson learned: the fastest approach is to use SQL exclusively (i.e. only use CLR to dynamically generate the SQL script), and to build the tree iteratively (not recursively). We create a cover index table which holds the completed levels of the tree, and for the next level we join the index table with the original table using the PARITION keyword to separate nodes, and ORDER BY and ROW_NUMBER() to find the median cut plane. The nodes are post-order numbered; this means that at query time, if an inner node does not need to be recursed further because its bounding box is contained in the query polyhedron, its child leaf nodes can be selected trivially using BETWEEN. According to theoretic runtime complexity considerations, kd-tree indexing performs optimally when the number of items in each leaf is equal to the number of leafs, which practically means that the number of leafs (and items in it) is equal to the square root of the number of rows. Thus our tree has 15 levels, $2^{14}$ leafs and in each leaf there are approximately 16K items. The run-time of the kd-tree generation over 270M rows was less than 12 hours.

When using the kd-tree for spatial queries, a CLR stored procedure recursively intersects the query polyhedron with the kd-tree bounding boxes (see Figure 4) until it finds nodes where further recursion is unnecessary or impossible (leaf node), and then performs a standard SQL query (as mentioned above) on the points in that node to get the final results.

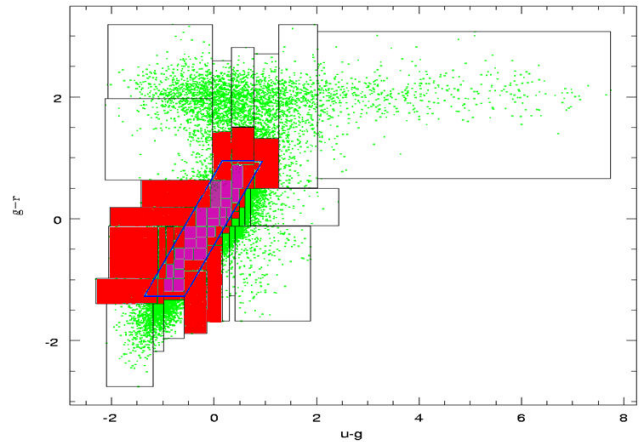

**Figure 4. A two dimensional demonstration of how a polyhedron query is evaluated at the leaf level of the kd-tree. Green dots are data points, empty cells are outside, purple cells are inside the query polyhedron. The database server has to perform an SQL query in the partially covered red cells.**

We have found that if the ratio of the returned / total number of rows is below 0.25 kd-trees can outperform simple SQL queries by orders of magnitudes.

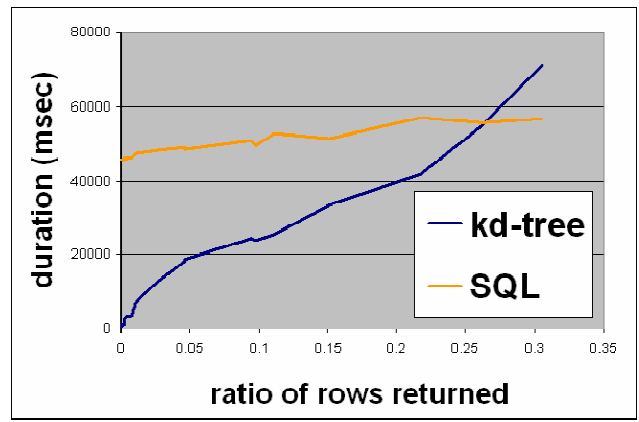

**Figure 5. Performance evaluation of the kd-tree index. For typical queries, where the number of returned points is a small fraction of the dataset, using the kd-tree index can speed up the query by orders of magnitudes.**

## 3.3 Kd-tree Based *k*-Nearest Neighbor Search

For scientific applications, to be discussed in Section 4, we needed a quick nearest neighbor procedure. The problem is simple: given a query point *p*, return the *k* nearest neighbors from the 270M magnitude table.

Most nearest neighbor search techniques employ two lists: the result list stores *k* points, the *k* nearest neighbors from the dataset we have examined so far. The other, the index list, stores some type of index, which identifies the set of points we have yet to examine in order to complete the program. As we examine more points from the index list we refine the *k* points in the result list, until the index list is empty, thus getting the *k* nearest neighbors.

This is the basic scheme of our technique, too. We find the leaf box in the kd-tree which contains *p*, and use a SELECT statement to select the *k* nearest neighbors from that leaf. The question at this point is which neighboring cells we have to examine to find the real nearest neighbors – the question of how to grow the index list. Note that our technique makes use of the fact that all boxes in the kd-tree are axis aligned. In order to find the indices of kd-boxes to examine, we use boundary points of kd-boxes we have already examined: a boundary point is either a vertex of a kd-box or the projection of *p* (along the three coordinates) onto the faces of the kd-boxes examined. Given *m*, the maximal distance of the points currently in the result list from *p*, the key idea is that if all boundary points are farther from *p* than *m*, then we are done. If *b* is a boundary point closer to *p*, add to the index list the kd-boxes not yet examined that include *b* as a boundary point, i.e. the kd-box(es) on the other side of *b*. Note that all points in this new kd-box *n* are farther from *p* than *distance(p, b)*, a fact we can use to refine our SQL on *n*: we store the result list in ascending distance order, and if for the first *f* points the distance from *p* is less than *distance(p, b)*, then there is no chance these will get replaced, i.e. only the remaining *k - f* are in danger, so we query *n* with *TOP(k - f)*. We insert these points into the result list if necessary and repeat the process until based on the distances of boundary points from *p* we find that no additional kd-boxes must be examined. The algorithm basically grows the region around p in steps of kd-boxes, maintaining a list of k-nearest neighbors, until it is impossible that points outside the grown region can replace the farthest point in the list, at which point the algorithm halts.

In practice we implemented a CLR stored procedure in C#, which takes the point *p* and *k* as inputs and returns the *k* nearest neighbors.

## 3.4 Voronoi Tessellation

Voronoi tessellation is a powerful technique for science data mining tasks: it is a natural method for similar object searches since it is an explicit solution of the nearest neighbor problem [6]. Additionally, because the volume of the cells is inversely proportional to the local density [1] (of data points) it can be used for finding clusters and outliers.

The Voronoi tessellation of a set of seed points defines a cell for each of the seed points in a way that all internal points of a cell are closer to the corresponding seed point than to any other seed points. The Voronoi cells are always convex polyhedra. The dual of Voronoi tessellation is called Delaunay triangulation; it is a graph connecting the seed points to its neighbors (i.e. two seed points are connected if their Voronoi cells share a common face). Compared to kd-trees, Voronoi tessellation is not a hierarchical structure, thus it is not appropriate to implement traditional logarithmic search algorithms without modifications. There are, however, techniques to overcome this limitation and build an index over the Voronoi cells [22] or use an approximate Voronoi diagram [6].

Parallel to our kd-tree indexing scheme, we experimented with creating an indexing scheme based on Voronoi tessellation. Compared to kd-trees, the major disadvantage of Voronoi tessellation is that it is much harder to compute (the complexity is $O(n\, logn)$ and non-polynomial if vertices are needed, too) and it takes much more space to store. Additionally, with kd-trees, there is no induced distance metric, whereas with Voronoi tessellation the distance between two points must be defined. We use the natural Euclidian metric; after whitening this should give correct results. On the other hand, Voronoi tessellation does have advantages: for our uneven distribution, standard kd-trees produce very elongated bounding boxes. Although there is a way to partly eliminate this effect in kd-trees [8] this problem usually does not arise with Voronoi tessellation, which produces nice, sphere-like polyhedra.

The memory requirements of Voronoi tessellation algorithms make it impossible to calculate the tessellation of our 270M magnitude table. Instead, we took a $N_{seed}$ = 10K representative sample; these were the seed points. We have chosen the seeds randomly, but this technique could be improved to follow better the underlying distribution, hence keep the cells balanced. We then used the open source QHull library [3] to compute its 5D Voronoi tessellation. QHull uses the Quick Hull method, which solves the dual problem of Voronoi tessellation, Delaunay triangulation, from which the former can be computed.

The cells were then numbered along a space filling curve and each of the 270M magnitude points were tagged with the ID of their enclosing Voronoi cell. A clustered index built over the tags allows us to retrieve all the points belonging to a specific Voronoi cell efficiently. To find the containing cell we used a directed walk on the Delaunay graph, which on average takes $O(sqrt(N_{seed}))$ steps. This index can be used to speed up polyhedron queries: for each of the $N_{seed}$ cells, we determine whether it is contained in the query or outside of it - in which case we return or reject, respectively, all points with that index -, or if it partially intersects, in which case we run the polyhedron SQL query. Note that detecting the intersection of two general polyhedra (the query and the Voronoi cell) is computationally a more challenging task than the same task for a rectangular cell and a polyhedron.

The ratio of the seed points vs. the total number of data points can be adjusted to find the optimum. Our expectation is that one of the main advantages of using Voronoi tessellation over rectangular boxing techniques is that with increasing dimensionality the shape of the cells will get more sphere-like than the shape of hyper-rectangles we would get for axis parallel cuts in a kd-tree. This is important for optimal covering (space filling) of uneven distributions. In practice, it turned out that Voronoi cells in five dimensions tend to have about a thousand vertices compared to the 32 for 5D hyper-rectangles and 50 neighboring cells ("faces") compared to 10 for hyper-rectangles. It confirms our expectation about the "roundness" of the cells.

In the future, we plan to compute the full Voronoi tessellation of the 270M magnitude table. Such a large partition cannot be computed in one pass, since all known algorithms would run out of memory (based on test runs, we estimate that the memory requirements for 270M points would be 270GB). The tessellation of such a large set would span several terabytes, storing it in the database is most likely not worth it. A possible solution is to store only the edges of the Delaunay triangulation, which is a much more compact description: we estimate that the Delaunay triangulation can be stored in 270GB. The obvious application of the Voronoi tessellation of the full 270M magnitude table is to use the inverse of the Voronoi cells' volume as a density estimator. This would give us a highly detailed, parameter-free density map of the entire magnitude space which may be used for redshift estimation and clustering.

## 3.5 Vector Data Type in SQL Server 2005

During the implementation of the above spatial indexing algorithms we intensively used SQL Server 2005's integrated CLR. In order to implement a unified data access layer to multidimensional databases a vector data type is essential. SQL Server's CLR User Defined Types (UDT) seemed to be a good choice to implement a vector data type, but it turned out that UDTs require a custom serializer in order to serialize C# arrays and use BinaryFormatter, which is much slower than native serialization. Performance tests showed that working with UDTs is a CPU consuming process, so we decided to use the simple binary data type and several unsafe C# functions to implement a vector-like data access method. Unsafe code (pointers and pointer arithmetic) is a useful feature of C# and helps in many cases when high performance code has to be mixed with the stability of managed features. We created a simple unsafe function that copied data from a SqlBinary variable to a typed array (float[] or double[]). The usage of unsafe code outperforms the UDTs in native serialization mode and it only slows down table scan queries by 20% compared to queries using only native SQL data types. This difference comes from the CLR function entry point calls.

## 4. SCIENTIFIC APPLICATIONS

We have created several stored procedures to exploit the capabilities of the spatial indices. We can quickly evaluate query polyhedron, find nearest neighbors or similar objects based on a training set and detect outliers based on the volume of the spatial bins. We outline a few case studies.

Unsupervised clustering and classification of objects may be performed using basin spanning trees (BST) [12]. The classification is based on the assumption that different spectral type objects are clustered in different regions of the color space (see Figure 1). We used the volumes of Voronoi cells to find density peaks (small cell volume means large local density), and connected each cell to one neighbor, the one with the largest density (see Figure 6.). Continuing this as a gradient process we separate density clusters. Comparing with the real classification for a subset where this information is available, we found that these clusters contain objects with the same spectral type (for 100K objects with a priori spectral classes 92% of objects were classified correctly).

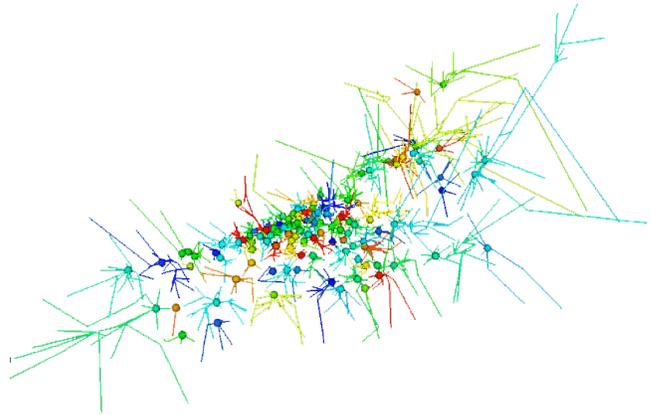

**Figure 6. The Basin Spanning Tree computed from a 100K sample of the magnitude table.**

In the next subsections we demonstrate the efficiency of the combination of spatial indexing and databases in two concrete astronomical applications. Both use our index-supported nearest neighbor finding procedure to quickly search the database.

## 4.1 Photometric Redshift Estimation

The first application is the so called *photometric redshift estimation* problem. Without going too deep into the astronomical details, the problem is the following: to answer basic astronomical and cosmological questions one would need a detailed three dimensional map of the Universe. From the telescope pointing direction and calibration to standard stars it is relatively easy to measure the two coordinates of an object on the sky. However, measuring the third spatial coordinate, the distance, is not this simple; it requires the exploitation of various physical laws. The first we exploit is that the Universe is expanding in an almost linear way, according to the so called Hubble law, i.e. the radial velocity of a galaxy is proportional to its distance from us. Thus if we could measure the velocity, we would have the third coordinate. Velocity cannot be measured directly either, but an effect similar to the Doppler shift (which makes the pitch of sound of an approaching and receding train higher and lower, respectively) works for light waves, too, making the light bluer or redder than the original color. In the expanding Universe all the galaxies are flying apart, so each should be redder than it would be in the rest frame. The amount of this *redshift* is related to the radial speed. To estimate this shift, astronomers spread the light of the galaxy with a prism, which gives the so called spectrum: the intensity of the light as a function of the wavelength. In this spectrum we can observe absorption and emission lines of well known chemical elements. These lines are shifted to redder wavelengths for receding objects compared to the standard values, the amount of this shift is the redshift. Measuring redshift is time consuming; SDSS spends 80% of its time to gather redshift for the brightest less than 1% of the objects. 5 colors without redshift for all the objects is gathered in the remaining 20% of the time. We have 5 colors for 270M galaxies and have redshifts for only 1M of them. It is not possible to draw an accurate map of the Universe if one coordinate is missing for 99% of the objects. Fortunately there is some – though not simple – relation between galaxy colors and redshift. In the last couple of years we have developed several 'data mining' methods which were able to harness this relation and estimate redshift from photometry. These *template fitting methods* are based on the convolution of template spectra and optical filter transmission curves. They require a substantial amount of computation and can only be run offline. The total computation on a 28 processor Blade server took almost 10 days to complete for the whole catalog. Another drawback of this technique is the difficulty in calibrating it to get rid of systematic observational errors.

With the help of our spatial indexing scheme for the color space and utilizing the database server's CLR support we have developed a non-parametric photometric redshift estimation method. The idea is to estimate redshifts based on a reference set. The *reference set* is the catalog of 1 million galaxies where both colors and redshifts were observed by the telescope. We will refer to the other set of the circa 270M objects with unknown redshifts as the *unknown set*. Since the reference set covers the color space relatively well, the redshift of a galaxy in the unknown set can be estimated with the redshift of the closest (with minimal Euclidean distance in color space) galaxies in the reference set. To smooth out the effects of the measurement errors we have found that instead of using the average, a local low order polynomial fit over the neighbors gives a better estimate. The pseudo code for the estimation is the following:

```
foreach (Galaxy g in UnknownSet)
{
  neighbors =
      NearestNeighbors(g, ReferenceSet)
  polynomCoeffs =
      FitPolynomial(neighbors.Colors,
                    neighbors.Redshifts)
  g.Redshift = Estimate(g.Colors,
                        polynomCoeffs)
}
```

In the first line of the loop, we utilize the nearest neighbor search method using our kd-tree index as described in Section 3.3.

The second and third line of the loop, the polynomial fit requires some intensive math calculation, including matrix inversion that would be prohibitive to do with native SQL. Also, pulling out the data from the server would impose a large time delay. Instead we have created a CLR stored procedure. This procedure uses a multi-parameter general least square fit code written in C# [16], compiled and run in the SQL Server utilizing the CLR support. Later we would like to replace the math routines with the AMD optimized LAPACK library [23]. The C# version runs reasonably fast to compute the estimation for the whole catalog; using the AMD Math Library should give ad additional speed up. Due to the fact that the nearest neighbor fitting method is not sensitive to calibration errors the precision of the estimation has also improved: average error decreased by more than 50%.

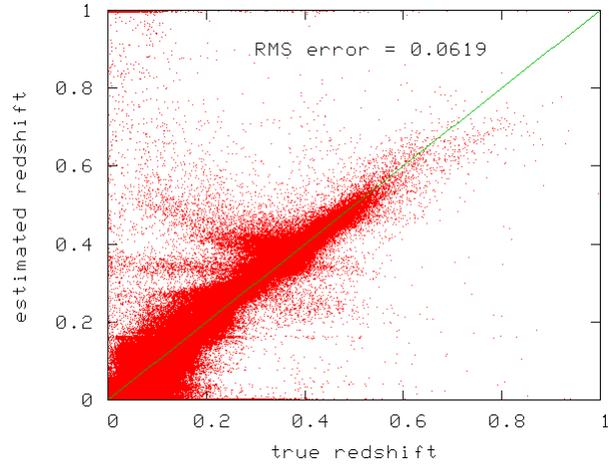

**Figure 7. Redshift estimation with the original template fitting method. Ideally, the estimated and true values should be equal, lying along the diagonal, but calibration problems of the templates produce large scatter.**

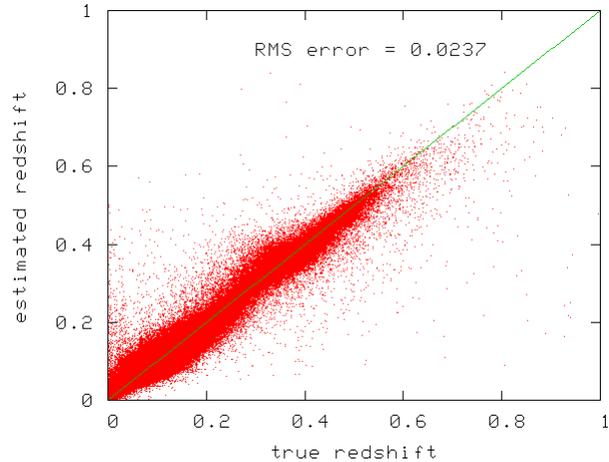

**Figure 8. Redshift estimation with k-nearest neighbor polynomial fit method. Estimation error is less than half compared to the template fitting method.**

Note that in theory the nearest neighbor search technique could have been implemented without a database, and would produce similar figures, but handling the large amount of the data with the database server simplifies much of the implementation. Also, since the SDSS database is accessible online through an SQL interface [27] by astronomers, they can use our stored procedure to estimate redshifts for new objects without downloading the reference set and running their own estimation code. This non-parametric estimation technique is not limited to astronomical application; e.g. we are currently working on porting it to a database containing network delay measurements.

### 4.2 Spectral Similarity Search
Our second application is spectral similarity search. In the previous section we introduced galaxy spectra. Galaxy spectra are not all the same, they depend on the age and composition of galaxies. Also, there are other objects like stars and quasars (point like active galaxy nuclei) in the catalog. Astronomers sometimes need to find similar objects to study their distribution in space, time evolution, etc. SDSS spectra are sampled at over 3000 wavelength values, so they are essentially 3000 dimensional vectors. They are stored in a separate archive, called SpectrumService [28]. Indexing of the 3000 dimensional space would be prohibitive, but it has been shown [9] that the first few principal components of the Karhunen-Loeve transform is enough to describe most of the physical characteristics. Essentially with a principal component transformation we can create a low (we have chosen 5) dimensional feature vector for galaxies. (Note that similar techniques are routinely used in image recognition methods.) Thus, for the spectral feature space a similar index can be built and the same stored procedures can be used for nearest neighbor searches as for the magnitude space. Figures 9. and 10. illustrate a successful search for similar galaxies and quasars.

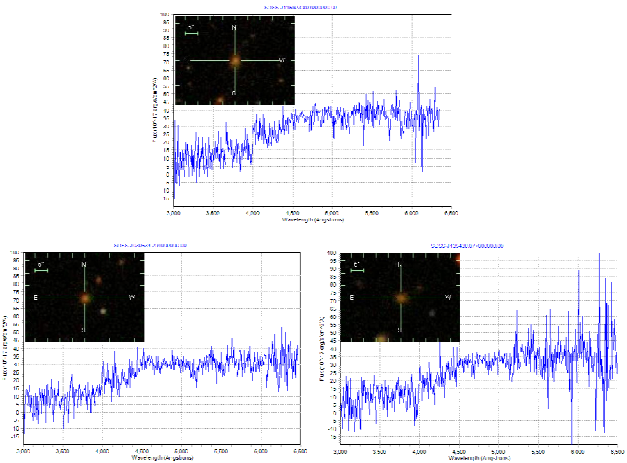

**Figure 9. The top figure is a typical elliptic galaxy and its spectrum. The lower two are the two most similar spectra based on the feature vector containing the first 5 principal components.**

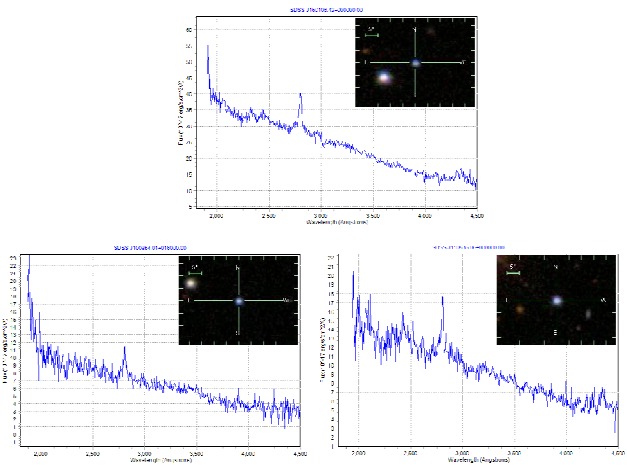

**Figure 10. The top figure is a typical quasar and its spectrum. The lower two are the two most similar spectra based on the feature vector containing the first 5 principal components.**

Note that we did not search for similar images; the similarity of the pictures of the celestial objects in Figures 9. and 10. are the result of similar spectra. We are working on feature extraction methods that would enable us to do similarity searches in the image space, too.

Observational archives are not the only science databases showing exponential growth: simulations also generate enormous data sets. Comparing simulations and observations is a challenging task. For example, there are simulations of galaxy spectral features: tweaking the age, chemical composition, dust content and other physical parameters, it is possible to generate spectra for various galaxy evolution scenarios. To decide which model is closest to reality, the simulated spectra should be compared with the observed ones. Our framework made it possible for us to do such a comparison between the 270M points of the SDSS data set and 100K spectra simulated by the Bruzual-Charlot spectral synthesis code [7]. The initial parameters of the closest spectra can be looked up, in this way astronomers can "reverse engineer" the observed data to estimate physical parameters of galaxies.

## 5. INTERACTIVE VISUALIZATION

Various tools exist for visualizing science data; a few of them even have some basic database connection interfaces. Initially, we have used KitWare's ParaView [25]. ParaView is an open-source application available for all major platforms. It uses the Visualization ToolKit (VTK) [26] at its core; in fact ParaView is but a thin GUI on top of VTK. The name ParaView is short for Parallel Visualization, since the application supports parallel visualization pipelines. ParaView has a powerful plugin architecture using which you can create your own custom data filters (transformers) and data sources.

The first version of our workflow was to save a dataset from our database to a flat text file; then write a C program which first reads the input data from the file, calculates some geometric structure in-memory and finally writes it to a file in ParaView's VTP format (see Figure 6). We would then launch ParaView to examine our work. Later we created a ParaView plugin which lets users enter SQL queries into a textbox and then transmits it to the database. The resulting three column table was interpreted as a 3D point set and visualized.

Problems arise, however, if your dataset is very large. Large can mean two things in this context: if the data is several gigabytes in size, the conversion step becomes slow or impossible due to system memory limits (e.g. the magnitude table of SDSS is 53GB). On the other hand, even if you surmount this problem, visualizing more than a few million objects is not possible on consumer-grade PCs, our target architecture.

The key idea is *adaptive visualization*: to choose the level of detail depending on where the user's virtual camera is. Adaptive visualization requires an event-driven architecture, where data sources can dynamically generate new datasets as the user changes his vantage point (e.g. zooms in). This was one of the key requirements that led us away from ParaView where this two-way interaction would be difficult to implement. Another was our desire for rapid prototyping. We wanted to leverage the powerful libraries and ease of use of some managed platform. Since our database end uses MS SQL Server 2005 and managed C# stored procedures, .NET was a natural choice. Additionally, with .NET we can use Managed DirectX for 3D visualization, whereas Java, the other platform we considered is typically not used for real-time 3D applications.

Writing our own visualization application from scratch has been a very rewarding project. Building on the standard .NET libraries and Managed DirectX, we arrived at an architecture that serves our visualizing needs quite well. The visualization program is only 5000 lines of code.

When designing the application, one of the major requirements was extensibility. We were inspired by the plugin architectures of some commercial applications as well as ParaView. The key difference between our plugin architecture and ParaView's is our support for input events such as camera movement. In our approach there are two types of plugins: producers and pipes. Producers are output-only objects; they are, from the visualization application's perspective, the source of all geometry data. The main visualization application is not aware of how each Producer plugin generates the geometry output it passes on to the pipeline connected to it, and eventually to the visualizer. In our case, all our Producers connect to the database to perform SQL queries whose result tables are interpreted by the code in the plugin; these results are used to create the 3D geometry data which is then passed to the visualizer pipeline (see Figure 11). Pipes are input/output objects which transform their input in some manner (they correspond to ParaView's filters). ParaView demonstrates that this is a very powerful paradigm: well designed pipes can be used in many visualization contexts.

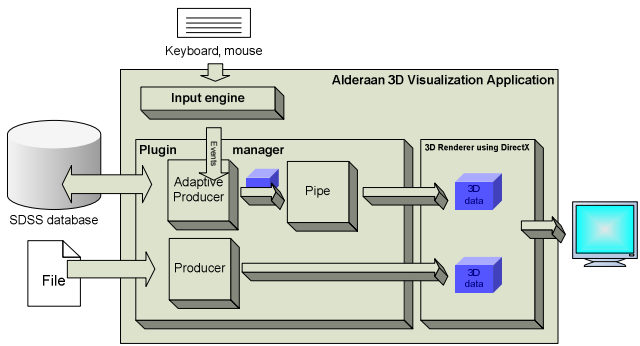

**Figure 11. The visualization pipeline of the adaptive visualization application. Producer plugins fetch data from the database upon receiving input events, generate and pass 3D geometry into the pipeline.**

## 5.1 Technical Details

In terms of software architecture, our application consists of three parts: the visualization application, the plugin interfaces and the actual plugins. The plugin interfaces contain the C# interfaces (Producer, Pipe) which the plugins must implement, and the definitions of data structures used to transfer 3D geometry data to and from plugins. This is all the plugins know about the application, and all that is needed to compile the plugins into dynamically linked libraries (DLLs). The Plugin interface is thus the sole connection point between a plugin and the application.

The architecture supports multi-threaded execution, i.e. it is set up in a way where plugins can run in a different thread. This is essential in cases where the plugin communicates with a database server to download data and then performs additional, possibly time-consuming operations to generate the output 3D geometry:

the main application must remain responsive. On the other hand, in simple cases, e.g. when a plugin loads data from a small file, a separate thread's programmatic complexity does not pay off. Thus our architecture is set up in a way to support both single-threaded and multi-threaded models.

For clarity, the important interfaces are reproduced below:

```
public interface Plugin
{
    bool Initialize(Registry registry);
    bool Start();
    bool Stop();
    void Shutdown();
}

public interface Producer : Plugin
{
    GeometrySet GetOutput();
    Camera SuggestInitial();
}
// interface definitions for Consumer and Pipe
// not shown

public interface Registry
{
    Event CameraChanged CameraBoxChanged;
    // other events not shown
    void SignalProduction(Producer producer);
}
```

**Figure 12. The Plugin interfaces.**

On startup, the application uses object metadata – Reflection in .NET – to find classes in DLL files implementing the Producer and Pipe interfaces. It then loads the configuration XML file, which contains the plugin graph. The appropriate plugins are then instantiated, each is passed a separate Registry object in the constructor, and Start() is called. Plugins use the Registry object to register to receive events such as camera change events. Upon receiving such an event, they compute the appropriate 3D geometry they would like to show. Once ready, they call their Registry's SignalProduction() function, whose semantic is to signal the main application that new data has been produced. The application will then call the plugin's GetOutput() function, which returns the 3D geometry in a GeometrySet object.

In the more interesting multi-threaded case, one must remember that functions such as GetOutput() and event delegates are called by the application, so they run in the main thread's context. They must take care not to perform blocking calls, which would render the main application unresponsive. However, one can only access the plugin's internal data structures after obtaining a lock, since the plugin thread might be writing it. Thus, the typical implementation of the GetOutput() function tries to obtain a lock using a non-blocking call, and if it fails, it returns null. In this case, the main application will attempt to extract the 3D geometry in the next frame cycle. On the other hand, the SignalProduction() function, which is implemented by the application, is called by the plugin, so it is executed in the plugin thread's context. In practice, this simply sets a flag to signal the application that in the next frame cycle it should attempt a GetOutput() call. The benefit of this mechanism is that the application does not have to wait for the plugin and vice versa, the plugin does not have to wait for the application to accept each production result. Practically, our plugins always store the last completed GeometrySet, so the only time GetOutput() will return null is when the plugin is replacing it with newer data.

The architecture allows plugins to cache their geometry. Our plugins save the last $n$ result sets, and when a camera change event is fired, they first look for geometry in this local, in-memory cache. The database is contacted only if additional geometry is needed. In practice, when zooming in and then back out, the cache reduces time delay to zero. Since this logic is entirely contained in the plugin, it does not increase our architecture's complexity. The threading mechanism is shown on Figure 13.

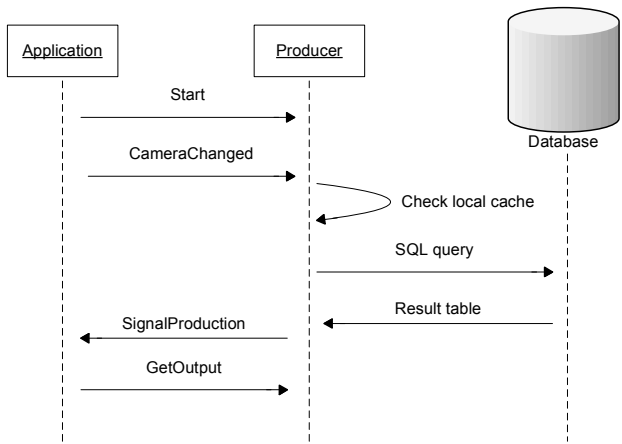

**Figure 13. Multi-threading plugin connecting to the database.**

## 5.2 Visualization Use-Cases

We use our application to explore the SDSS dataset and to visualize the spatial indexing schemes. Each visualization corresponds to a Producer plugin which contacts the database to retrieve rows from pre-defined tables or stored procedures and locally caches results. Currently, the plugins fall into three categories: ones which adaptively visualize point clouds, ones which adaptively visualize kd-trees, and ones adaptively visualizing Delaunay and Voronoi structures built above the 270M magnitude table.

We use point data to visualize the first three principal components of the magnitude table of the SDSS database. Our Producer add-on responds to camera changes by first checking its local cache, and if necessary querying the server for new points to ensure that there are at least $n$ (we use $n = 100K$) objects in view. Our other point cloud visualization is that of the SDSS *ra*, *dec*, *redshift* space. In astronomy, right ascension (*ra*) and declination (*dec*) are the two standard angles to parameterize a unit sphere. Using Hubble's law, which states that celestial objects farther away are receding faster and thus have higher redshift (and these relations are linear), we can trivially compute the radial distance of celestial objects from redshift data. This visualization thus shows the 3D spatial distribution of the celestial objects measured by the SDSS telescope, as seen from the Earth. This shows the large scale structure of the universe (e.g. Finger of God structures) in an adaptive manner (see Figure 14).

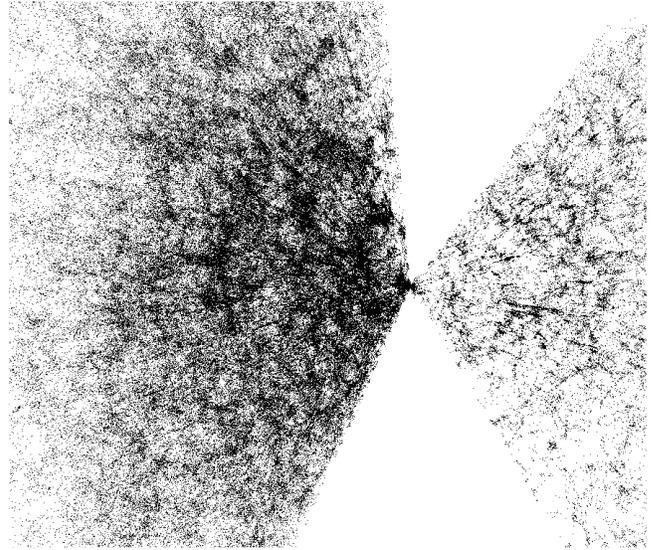

**Figure 14. Visualization of the large scale structure of the Universe. The visualization client is displaying 500K points every frame. Each point represents a galaxy, and additional structure, clusters of galaxies are clearly visible. Adaptive visualization allows astronomers to zoom in and explore individual clusters in 3D, similar to browsing satellite maps with Google Earth or NASA's World Wind.**

We created a Producer plugin which queries the kd-tree of the 270M magnitude table and displays the sub-tree according to the visualization camera at an appropriate depth so that at least $n$ (we use $n = 500$) kd-boxes are visible. We created a separate plugin which also queries the point table, and shows points and kd-boxes together (Figure 15).

Our Delaunay graphs and Voronoi tessellations used for visualizing were not created using the full 270M SDSS magnitude table. We are still working on a method that is able to compute the Delaunay graph of such a large dataset. However, in terms of visualization, our application is up to the task. As a demonstration, we exported a 1K, a 10K and 100K sample of the magnitude table and computed its Delaunay graph in-memory and imported it back into the database. This enables us to do a 3-level adaptive visualization of the 100K sample. The plugins query the Delaunay graph of the 1K point table, and if not enough edges are returned, it goes on to the 10K and subsequently 100K tables to ensure a good level of detail. We have two plugins: the Delaunay plugin simply displays the edges returned; the Voronoi plugin uses the edges returned and computes and displays the induced Voronoi-cells (Figure 16).

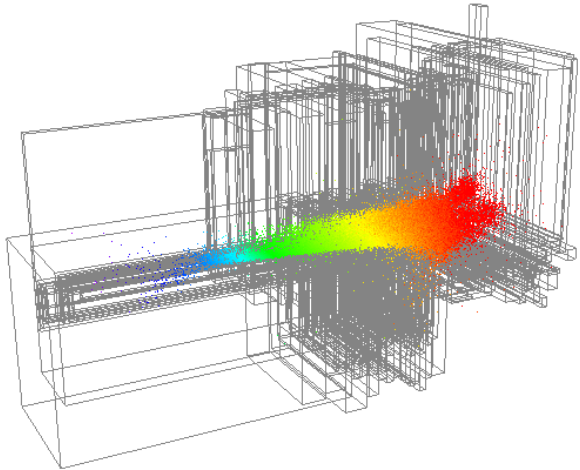

**Figure 15. The kd-tree used for indexing and the underlying point distribution. Boxes tend to be elongated along the second and third principal components.**

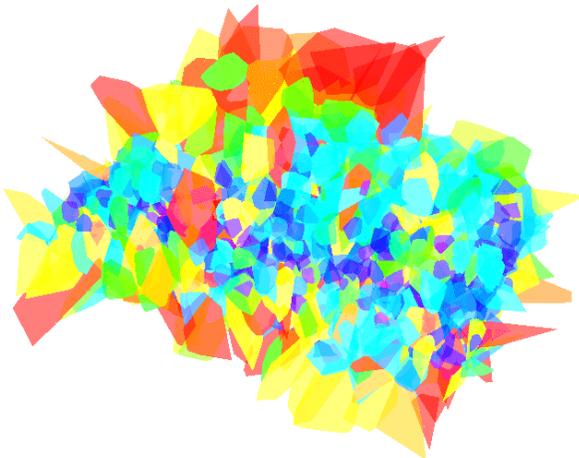

**Figure 16. Visualization of the adaptive Voronoi tessellation. Colors correspond to the volume of cells.**

## 5.3 Lessons Learned

Visualizing large datasets, in terms of software engineering, poses two tasks: creating the appropriate indexes and support libraries (as stored procedures) on the database server side and writing the visualization application on the client side. Our experience shows that the latter is a task where only minor technical innovations are required, due to the sheer amount of experience the field has accumulated from 3D games. The former, on the other hand, does require innovation: non-trivial challenges arise when creating spatial indices on large datasets as well as when attempting to quickly query these.

The real challenge in visualization is not a technical one. It is to create a workflow that enables the scientist to do something *new and useful* he was not able to do before. Thus, our long term plans are to find visualization use-cases that truly benefits the SDSS astronomer by helping him make novel new discoveries.

## 6. CONCLUSION

We have implemented a prototype system that is able to efficiently index large amounts of multidimensional continuous science data. With the aid of spatial indices it can quickly execute complex geometric queries and be used for interactive, adaptive data visualization. We have demonstrated the system and its applicability on an astronomical archive, but we emphasize that our database is not unique. Integration of managed procedural programs into the database engine (e.g. CLR in MS SQL Server 2005) decreased development time and complexity.

Future plans involve the implementation of further indices and stored procedures for spatial searches and further development of the adaptive visualization tool. The completed basic framework effectively decreases the complexity of creating index algorithms and visualization plugins for new science applications.

## 7. ACKNOWLEDGMENTS

This work was partly supported by: OTKA-T047244, MSRC-2005-038, NAP-2005/KCKHA005 and MRTN-CT-2004-503929.

## 8. REFERENCES

[1] Y. Ascasibar and J. Binney "Numerical estimation of densities" *Monthly Notices of Royal Astronomical Society, Vol. 356, 872-882* (2005)

[2] F. Aurenhammer "Voronoi Diagrams —A Survey of a Fundamental Geometric Data Structure" *ACM Computing Surveys, Vol. 23, No. 3,* 345 - 405 (1991)

[3] C.B. Barber, D.P. Dobkin, and H.T. Huhdanpaa, "The Quickhull algorithm for convex hulls," *ACM Trans. on Mathematical Software*, 22(4):469-483, (1996) http://www.qhull.org

[4] J. S. Beis and D. G. Lowe "Shape indexing using approximate nearestneighbour search in high-dimensional spaces" *in Conference on Computer Vision and Pattern Recognition* (1997)

[5] J. L. Bentley. Multidimensional binary search trees used for associative searching. *Commun. ACM*, 18(9):509{517, (1975)

[6] S. Berchtold, D.A. Keim, H. Kriegel and T. Seidl "Indexing the Solution Space: A New Technique for Nearest Neighbor Search in High-Dimensional Space", *IEEE Transactions on Knowledge and Data Engineering, Vol.12, 1,* 45-57 (2000)

[7] G. Bruzual and S. Charlot "Stellar population synthesis at the resolution of 2003", *Monthly Notices of the Royal Astronomical Society, Volume 344, Issue 4,* 1000-1028 (2003)

[8] A. Chaudhary, A. S. Szalay, and A. W. Moore "Very fast outlier detection in large multidi-mensional data sets" *In Proc. ACM SIGMOD Workshop on Research Issues in Data Miningand Knowledge Discovery (DMKD),* (2002)

[9] Connolly, A. J., Csabai, I., Szalay, A. S., Koo, D. C., Kron, R. G., & Munn, J. A. "Slicing Through Multicolor Space: Galaxy Redshifts from Broadband Photometry" *Astronomical Journal 110 2655C* (1995)

[10] I. Csabai, A.S. Szalay, R. Brunner and K. Ramaiyer "Multidimensional Index for Highly Clustered Data with

Large Density Contrasts" *in Statistical Challenges in Modern Astronomy II, Pennsylvania State University, Springer-Verlag,* p.447 (1997)

[11] J. Gray et al. "There Goes the Neighborhood: Relational Algebra for Spatial Data Search" *Microsoft Research Technical Report* 32 (2004)

[12] F.A. Hamprecht et al. "A strategy for analysis of molecular equilibrium simulations: Configuration space density estimation, clustering, and visualization" *Journal of Chemical Physics Vol. 114, No.5* 2079-2089 (2001)

[13] G. Heber, C. Pelkie, A. Dolgert, J. Gray and D. Thompson "Supporting Finite Element Analysis with a Relational Database Backend; Part III: OpenDX – Where the Numbers Come Alive" *Microsoft Research Technical Report* 151 (2005)

[14] J. Kubica and A. Moore "Variable kd-tree algorithms for spatial pattern search and discovery" *in Advances in Neural Information Processing Systems,* (December 2005)

[15] M. Nieto-Santisteban, J. Gray, A.S. Szalay, J. Annis, A. Thakar, W. O_Mullane "When Database Systems Meet the Grid", *Proceedings of the 2005 CIDR Conference*, 154-161 (2005).

[16] W.H. Press, B.P. Flannery, S.A. Teukolsky and W.T. Vetterling "Numerical Recipes: The Art of Scientific Computing" *Cambridge University Press,* (1992)

[17] C. Stoughton, R. Lupton et al. "Sloan Digital Sky Survey: Early Data Release", *The Astronomical Journal, 123*, 485-548 (2002), see also http://www.sdss.org

[18] A. Szalay and J. Gray "Science in an Exponential World", *Nature 440* 413-414 (2006).

[19] A. Szalay, J. Gray, G. Fekete, P. Kunszt, P. Kukol, A. Thakar "Indexing the Sphere with the Hierarchical Triangular Mesh", *Microsoft Research Technical Report* 123 (2005)

[20] F. Takens "Detecting strange attractors in turbulence", *Lecture notes in mathematics, Vol.898. Dynamical systems and turbulence, Springer, Berlin*, p366, (1981)

[21] D. Watson "Computing the n-dimensional Delaunay tessellation with applications to Voronoi polytopes", *The Computer Journal, 24(2):167-172, 1981*

[22] H. Samet "Foundations of Multidimensional and Metric Data Structures", *Morgan Kaufmann Publishers*, 2006

[23] http://developer.amd.com/acml.jsp

[24] http://www.ivoa.net

[25] http://www.paraview.org

[26] http://public.kitware.com/VTK/

[27] http://skyserver.sdss.org

[28] http://voservices.net/spectrum